\documentclass[preprint,12pt]{elsarticle}

\usepackage{amssymb}

\journal{Information Processing Letters}

\usepackage{amssymb,amsmath}
\usepackage[ruled,vlined]{algorithm2e}

\newcommand{\commentOut}[1]{}

\newtheorem{theorem}{Theorem}
\newtheorem{question}{Question}

\begin{document}

\begin{frontmatter}



\title{On pattern matching with $k$ mismatches and few don't cares}


\author[uconn]{Marius Nicolae\corref{cor1}}
\ead{marius.nicolae@engr.uconn.edu}

\author[uconn]{Sanguthevar Rajasekaran}
\ead{rajasek@engr.uconn.edu}

\cortext[cor1]{Corresponding author}

\address[uconn]{Department of Computer Science and Engineering,
University of Connecticut, 371 Fairfield Way Unit 4155, Storrs, CT 06269, USA}

\begin{abstract}
We consider the problem of pattern matching with $k$ mismatches, where there can
be don't care or wild card characters in the pattern. Specifically, given a
pattern $P$ of length $m$ and a text $T$ of length $n$, we want to find all occurrences of $P$
in $T$ that have no more than $k$ mismatches. The pattern can have don't
care characters, which match any character. Without don't cares, the best known
algorithm for pattern matching with $k$ mismatches has a runtime of $O(n\sqrt{k
\log k})$. With don't cares in the pattern, the best deterministic algorithm
has a runtime of $O(nk \;\text{polylog}\; m)$.
Therefore, there is an important gap between the versions with and without don't cares. 

In this paper we give an algorithm whose runtime  increases with the number of
don't cares.
We define an {\em island} to be a maximal length substring of $P$ that does not
contain don't cares. Let $q$ be the number of islands in $P$. We present an
algorithm that runs in $O(n\sqrt{k\log m}+n\min\{\sqrt[3]{qk\log^2
m},\sqrt{q\log m}\})$ time. If the number of islands $q$ is $O(k)$ this runtime
becomes $O(n\sqrt{k\log m})$, which essentially matches the best known runtime
for pattern matching with $k$ mismatches without don't cares. If the number of
islands $q$ is $O(k^2)$, this algorithm is asymptotically faster than the
previous best algorithm for pattern matching with $k$ mismatches with don't
cares in the pattern.
\end{abstract}


\end{frontmatter}


\section{Introduction}

The problem of string matching can be defined as follows. Given a text $T=t_1
t_2\cdots t_n$ and a pattern $P=p_1p_2\cdots p_m$, with letters from an alphabet $\Sigma$,
find all the occurrences of the pattern in the text.
This problem can be solved in $O(n+m)$ time by using well known algorithms
(e.g., KMP \cite{KMP77}). 

A more general formulation allows ``don't care'' or ``wild card''
characters in the text and/or the pattern. Pattern matching with don't cares can
be solved in $O(n \log |\Sigma| \log m)$ as shown in \cite{FP74}. A more
recent result \cite{CC07} gives a deterministic $O(n \log m)$ time algorithm.

Yet another enhancement is to allow for mismatches. We can
formulate two versions of this problem: {\bf 1) pattern matching with mismatches:} 
 find the distance between the pattern and the text for every alignment
between the pattern and the text or {\bf 2) pattern matching with $k$
mismatches:} find only alignments for which the distance is no
more than a given threshold $k$.

The distance metric used can be the Hamming distance, the edit distance or
other criteria such as the number of non-overlapping inversions (e.g.
\cite{CC+13}). In this paper we focus on the Hamming distance.
The Hamming distance between two strings $A$ and $B$ is defined as the number
of positions where the two strings differ and is denoted by $Hd(A,B)$. 

Pattern matching with mismatches can be solved, naively, by computing the
Hamming distance for every alignment of the pattern in the text, in time $O(nm)$. However, the fastest known exact algorithm is
Abrahamson's algorithm \cite{ABR87} that runs in $O(n \sqrt{m \log m})$
time.

Pattern matching with $k$ mismatches can be solved in $O(nk)$
time (see \cite{LV85} and \cite{GG86}). These algorithms are based on a
technique called the Kangaroo method (see section \ref{sec_kangaroo}). This
method computes the Hamming distance for every alignment in $O(k)$ time by
``jumping'' from one error to the next. A faster algorithm for pattern
matching with $k$ mismatches runs in $O(n\sqrt{k \log k})$ \cite{ALP04}. A
simpler version of this algorithm was given in \cite{NR15}.

Recent work has also addressed the online
version of pattern matching, where the text is received in a
streaming model, one character at a time, and it cannot be stored in its
entirety (see e.g., \cite{CKP08}, \cite{PP09}, \cite{PL07}).
Another version of this problem matches the pattern against multiple input
streams (see e.g., \cite{CEP+07}). Yet another interesting problem is to sample a
representative set of mismatches for every alignment (see e.g., \cite{CEP+12}).
A survey of string matching with mismatches is given in \cite{Nav01}.
A description of practical on-line string searching algorithms can be
found in \cite{NR02}.

Yet another formulation allows for don't care or wild card characters.
Pattern matching with mismatches and don't cares can be solved in $O(n \sqrt{g \log m})$ time, where $g$ is the number of non-wild card
positions in the pattern (see \cite{NR15}). This is done by a simple extension of Abrahamson's algorithm.

Pattern matching with $k$ mismatches and don't cares can be solved in time
$O(nk^2\log^2m)$ as shown in \cite{Clif10}. The runtime can be improved to 
 $O(nk\;\text{polylog}m)$ as shown in \cite{Clif10, CEP+09}
If we allow don't cares only in the pattern, the problem can be solved in
$O(n\sqrt[3]{mk\log^2m})$ time as shown in \cite{CP07}. 
This is also the problem we discuss in this paper.

{\bf Notation:} Let $T_i$ denote $t_i t_{i+1},\ldots t_{i+m-1}$ for all
$i=1..n-m+1$.
 
{\bf Pattern matching with $k$ mismatches and don't cares in the pattern:}
Given a text $T=t_1t_2\ldots t_n$ and a pattern $P=p_1p_2\ldots p_m$ from an
alphabet $\Sigma$, with $|\Sigma|\leq n$, and an integer $k$.  Output all $i$, $1\leq
i\leq n-m+1$, for which $Hd(P, T_i) \leq k$.
The pattern may contain don't care characters, that match any character.

Given a pattern $P$, with don't cares, a maximal length substring of $P$ that
has no don't cares is called an ``{\bf island}''. We will denote the number of
islands in $P$ as $q$.
In this paper we give two algorithms for pattern matching with $k$ mismatches
where there are don't cares in the pattern. The first one runs in
$O(n\sqrt{(q+k)\log m})$ time. The second one runs in time $O(n\sqrt[3]{qk\log^2 m} +
n\sqrt{k\log m})$ where $q$ is the number of islands in $P$. By combining the
two, we show that pattern matching with $k$ mismatches and don't cares in the
pattern can be solved in $O(n\sqrt{k\log m}+n\min\{\sqrt[3]{qk\log^2 m},\sqrt{q\log m}\})$ time.
If the number of islands is $O(k)$ our runtime becomes $O(n\sqrt{k \log m})$,
which essentially matches the best known runtime for pattern matching with $k$
mismatches without don't cares ($O(n\sqrt{k\log k})$). Since $q$ is always less
than $m$, our algorithm outperforms the $O(n\sqrt[3]{mk\log^2m})$ algorithm of
\cite{CP07}.
For  $q=O(k^2)$, our algorithm outperforms the best known $O(nk
\;\text{polylog}\; m)$ algorithms of \cite{Clif10, CEP+09}.

\section{Methods}

Both algorithms in this paper have the same basic structure (see section
\ref{sec_basic}).
The difference is in how fast we can answer the single alignment verification question:

\begin{question}
Given $i$, is the Hamming distance between $P$ and $T_i$ no more than
$k$?
\end{question}

In the first algorithm (section \ref{sec_alg1}), we can answer this question in
$O(q+k)$ time. In the second algorithm (section \ref{sec_alg2}), we can answer
this question in $O(\sqrt[3]{k^2q^2\log m} + k)$ time.

\subsection{Background}

We start by reviewing a number of well known techniques
used in the literature for pattern pattern matching with $k$ mismatches (e.g., see
\cite{ALP04}), namely:
convolution, marking, filtering and the Kangaroo method.

\subsubsection{Convolution}
Given two arrays $T=t_1t_2\ldots t_n$ and $P=p_1p_2\ldots
p_m$ (with $m\leq n$), the convolution of $T$ and $P$ is a sequence
$C=c_1,c_2,\ldots,c_{n-m+1}$ where $c_i=\sum_{j=1}^mt_{i+j-1}p_j$, for $1\leq
i\leq (n-m+1)$.

Convolution can be applied to pattern matching with mismatches, as follows.
Given a string $S$ and a character $\alpha$ define string $S^{\alpha}$
as $S^\alpha[i]=1$ if $S[i]=\alpha$ and $0$ otherwise.
Let $C^\alpha=convolution(T^\alpha, P^\alpha)$. Then $C^\alpha[i]$ gives the
number of matches between $P$ and $T_i$
where the matching character is $\alpha$. Therefore, one convolution gives us
the number of matches contributed by a single character to each of the
alignments. Then $\sum_{\alpha \in \Sigma}C^{\alpha}[i]$ is the total number of
matches between $P$ and $T_i$.

One convolution can be computed in
$O(n\log m)$ time by using the Fast Fourier Transform. 
If the convolutions are applied on binary inputs, as is often the case in
pattern matching applications, some speedup techniques are presented in \cite{FG09}.

\subsubsection{Marking}\label{sec_marking}

Marking is an algorithm that counts the number of matches of
every alignment, as follows.
The algorithm scans the text one character at a time
and ``marks'' all the alignments that would produce a match between the current
character in the text and the corresponding character in the pattern. 
The marking algorithm is generally used only on a subset of the pattern. That
is, given a set $A$ of positions in $P$ the marking algorithm counts matches
between the text and the subset of $P$ given by $A$. The pseudocode of the
marking algorithm is given in Algorithm \ref{alg_counting}.

\begin{algorithm}
\SetKwInOut{Input}{input}\SetKwInOut{Output}{output}
\caption{Mark$(T, P, A)$}\label{alg_counting} 
\Input{Text $T$, pattern $P$ and a set $A$ of positions in $P$} 
\Output{An array $M$ where $M[i]$ gives the number of matches between $T_i$
and $P$, on the subset of positions of $P$ given by $A$}
\lFor {$i\leftarrow 1$ \KwTo $n$}{$M[i]=0$}
\For {$i\leftarrow 1$ \KwTo $n$}{
  \For{$j \in A$ s.t. $P[j] = T[i]$}{
    \lIf{$i-j+1 > 0$}{
      $M[i-j+1]${\bf $++$}
    }
  }
}
\Return $M$\;
\end{algorithm}

\subsubsection{Filtering}

Filtering is a method for reducing the number of alignments to look
at. Filtering is based on the following principle.
If we restrict our pattern to only $2k$ positions, any alignment that has no more than
$k$ mismatches, must have at least $k$ matches among the $2k$ positions.
To count matches among the $2k$ positions selected, for every alignment, we use
the marking algorithm. If the total number of marks generated is $B$ then there can
be no more than $B/k$ positions that have at least $k$ marks. Therefore, instead
of $n-m+1$ alignments we only have to look at $B/k$ alignments. Each alignment
is then verified using other methods.

\subsubsection{The Kangaroo method}\label{sec_kangaroo} 

The Kangaroo method allows us to check if the number of mismatches for a
particular alignment is no more than $k$, in $O(k)$ time. The Kangaroo method
constructs a generalized suffix tree of $T+P$, where $+$ means concatenation.
This suffix tree can be enhanced to answer Lowest Common Ancestor
(LCA) queries in $O(1)$ time \cite{AH+76}. LCA queries give us the longest
common prefix between any portion of the text and any portion of the pattern,
essentially telling us where the first mismatch appears. 
Specifically, to count mismatches between
$P$ and $T_i$, first perform an LCA query to find the position of the
first mismatch between $P$ and $T_i$. Let this position be $j$. Then,
perform another LCA to find the first mismatch between $P_{j+1..m}$ and
$T_{i+j+1.. i+m-1}$, which gives the second mismatch of alignment $i$.
Continue to ``jump'' from one mismatch to the next, until the end
of the pattern is reached or we have found more than $k$ mismatches.
Therefore, after $O(k)$ LCA queries we will either find all the mismatches or
determine that there are more than $k$ of them. 
The Kangaroo pseudocode is given in Algorithm \ref{alg_kangaroo}.

\begin{algorithm}
\SetKw{LCA}{LCA}
\SetKw{True}{true}
\SetKw{False}{false}
\SetKwInOut{Input}{input}\SetKwInOut{Output}{output}
\caption{Kangaroo$(P, T_i, k)$}\label{alg_kangaroo}
\Input{A pattern $P$, an alignment $T_i$ and an integer $k$}
\Output{\True if the pattern matches the alignment with no more than $k$
mismatches, \False otherwise}
$j=0$\;
$d=0$\;
\While{$d \leq k$} {
  $j = j + \LCA(T_{i+j}, P_{j+1})+1$\;
  \If {$j > m$} {
     \Return{\True}\;
  }
  $d=d+1;$
}  
\Return{\False}\;
\end{algorithm}

\subsection{General Algorithm}\label{sec_basic}

We are now ready to present the main algorithms given in this paper. The
general structure of both the algorithms is given in Algorithm \ref{alg_basic}.

\begin{algorithm}
\caption{$K$-Mismatches with Wild Cards}\label{alg_basic}
Let $F_a$ be the number of occurrences of character $a$ in $T$ for all $a \in
\Sigma$\; 
Let $Cost(A)=\Sigma_{i \in A}F_{P[i]}$\;
Let $A$ be a set of positions in $P$ such that $|A|\leq 2k$
and $Cost(A) \leq B$\; 
$M=Mark(T, P, A)$\;
\eIf{$|A| == 2k$}{
   $R=\{\}$\;
   \For{$i=1$ to $n$} {
      \If {$M_i \geq k$ {\bf and} $DistNoMoreThanK(T_i, P, k)$} {
          $R = R \cup \{i\}$\;
      }
   }
}{
   
   \For{$a \in \Sigma$ s.t. $a \neq P[i], \forall i \in A$} {
      $M'=Convolution(T,P,a)$\;
      $M \text{+=} M'$\;
   }
   $R = \{i \in [1..n] | M_i \geq m - k\}$\;  
}
\Return{$R$}\;
\end{algorithm}

{\bf Algorithm and analysis:} 
For each position $i$ in $P$ such that $P[i]=a$, we assign a cost $F_a$ where
$F_a$ is the number of occurrences of $a$ in $T$. The algorithm starts by
choosing up to $2k$ positions from the pattern such that the total cost does not exceed a ``budget''
$B$. The positions are chosen by a simple greedy strategy: sort all the
characters by their cost $F_a$. Start choosing positions equal to the ``cheapest''
character, then choose positions equal to the next cheapest character, and
so on until we have chosen $2k$ positions or we have exceeded the budget $B$.

{\bf Case 1:} If we can find $2k$ positions that cost no more than $B$, then
we call the marking algorithm with those $2k$ positions. Any
position in $T$ that receives less than $k$ marks, has more than $k$ mismatches,
so we now focus on positions in $T$ that have at least $k$ marks.
If the total number of marks is $B$, then there will be no more than
$B/k$ positions that have at least $k$ marks. We verify each of these
positions to see if they have more than $k$ mismatches. Let the time for a
single verification be $O(V)$.
Then, the runtime is $O(BV/k)$.

{\bf Case 2:} If we cannot find $2k$ positions that cost no more than $B$,
then we compute marking for the positions that we did choose before we ran out
of budget.
Then, for each of the characters that we did not choose, we compute one
convolution to count how many matches they contribute to each alignment. It
is easy to see that each of the characters not chosen for marking must have $F_a
> B/(2k)$.
Therefore, the total number of such characters is no more than $n/(B/(2k))$. Therefore, the runtime of the convolution stage is $O(nk/B * n \log m)$. The runtime of the marking
stage is $O(B)$, therefore the total runtime is $O(B + nk/B * n \log m)$.

If we make the runtime of the two cases equal, we can find the optimal value of
$B$.

\begin{align*}
BV/k = B+n^2k/B \log m \Rightarrow B=nk\sqrt{\frac{\log m}{V}}
\end{align*}

This gives an asymptotic runtime of $O(BV/k)=O(n\sqrt{V \log m})$. Therefore,
the runtime of the algorithm depends on $V$, which is the time it takes to
verify whether a given single alignment has no more than $k$ mismatches.
 
\subsection{Single alignment distance in $O(q+k)$ time}
\label{sec_alg1}

We can answer the single alignment question 
in $O(q+k)$ time where $q$ is the number of {\it islands} in the pattern as
shown in Algorithm \ref{alg_verif1}.
The algorithm uses Kangaroo jumps \cite{LV85} to go to the next mismatch within
an island in $O(1)$ time. If there is no mismatch left in the island, the algorithm goes
to the next island also in $O(1)$ time.
Therefore, the runtime is $O(q+k)$. With $V=O(q+k)$, Algorithm \ref{alg_basic} does pattern matching
with $k$ mismatches in $O(n\sqrt{(q+k)\log m})$ time.

\begin{algorithm}
\label{alg_verif1}
\caption{$DistNoMoreThanK\_V1(T_i, P, k)$}
$d=0$\;
$j=1$\;
\While{$d \leq k$ {\bf and} $j \leq q$}{
  $r =$ no. of mismatches between island $j$ and
  corresponding region of $T_i$ (use Kangaroo jumps)\; 
  $d \text{+=} r$\; 
  $j \text{+=} 1$\;   
}
\Return{$d \leq k$}
\end{algorithm}

\subsection{Single alignment distance in $O(k^{2/3}q^{2/3}\log^{1/3}m+k)$ time}
\label{sec_alg2}

This idea is based on splitting the pattern into sections. We know that no more
than $k$ sections can have mismatches. The remaining sections have to match
exactly. Consider exact pattern matching with don't cares.
We can check where a pattern matches the text exactly by using a constant number of convolutions. This is
true because we can compute the values $C_i = \Sigma_{j=0}^{m-1}(T_{i+j}-P_j)^2T_{i+j}P_j$ using a constant
number of convolutions (see  \cite{CC07}). If $C_i=0$ then the pattern matches
the text at position $i$. 

Using this result, we will split the pattern into $S$
sections. In each section we include $q/S$ islands. For each of the $S$
sections, we use a constant number of convolutions to check where the section
matches the text. If $P$ has no more than $k$ mismatches at a particular
alignment, then at least $S-k$ sections have to match exactly. Each of the at
most $k$ sections that do not match exactly are verified using Kangaroo jumps as seen
earlier. One section takes at most $O(q/S+k')$ time, where $k'$ is the number
of mismatches discovered in that section. Over all the sections, the $k'$
terms add up to no more than $k$, therefore the entire alignment can be verified
in time $O(S+k+kq/S)$.

If we make $V=O(S + k + kq/S)$ in Algorithm \ref{alg_basic}, then its runtime
becomes $O(n \sqrt{V\log m}) = O(n \sqrt{(S + k + kq/S)\log m})$. 
The preprocessing time for the $S$ sections is $O(Sn \log m)$. The
optimal value of $S$ is such that the preprocessing equals the main runtime:

\begin{align*}
 & n \sqrt{(S + k + kq/S)\log m} = Sn \log m \\
\Rightarrow & S + k + kq/S = S^2 \log m\\
\Rightarrow & S^2/\log m + kS/\log m + kq/\log m = S^3\\
\Rightarrow & S \approx O(\sqrt[3]{kq/\log m})
\end{align*}

This makes $V=O(S + k + kq/S)=
O(k +\sqrt[3]{k^2q^2\log m})$. This gives a
runtime for pattern matching with $k$ mismatches of:

\begin{align*}
O(nS\log m + n \sqrt{V\log m}) = & O\left(n\sqrt[3]{kq\log^2
m} + n\sqrt{ (k + \sqrt[3]{k^2q^2\log m}) \log m   }\right)\\
= & O\left( n \sqrt[3]{kq\log^2 m} +  n\sqrt{k \log m} \right)\\
\end{align*}

\subsection{Combined result}
If $q < k^2$ then we can use the algorithm of section
\ref{sec_alg1}, which runs in $O(n\sqrt{(q+k)\log m})$ time. Otherwise, if
$q > k^2$, we use the algorithm of section \ref{sec_alg2}, which  runs in
$O(n\sqrt[3]{qk\log^2 m} + n\sqrt{k \log m})$ time.
Thus we have the following:

\begin{theorem}
Pattern matching with $k$ mismatches, with don't care
symbols in the pattern, can be solved in
$O\left(n\sqrt{k \log m} + n\min\{\sqrt{q\log m}, \sqrt[3]{qk\log^2
m}\}\right)$ time.
\end{theorem}

\section{Conclusions}
In this paper we have offered efficient algorithms for the problem of pattern matching with $k$ mismatches. Specifically,
we have presented an algorithm that runs in
$O(n\sqrt{k\log m}+n\min\{\sqrt[3]{qk\log^2 m},\sqrt{q\log m}\})$ time, where $q$ is the number of islands. If the number of islands $q$ is $o(m)$, this algorithm is asymptotically
faster than the previous best algorithm for pattern matching with $k$ mismatches
with don't cares in the pattern.

\section{Acknowledgments}
This work has been supported in part by the following grants: NSF 1447711 and NIH R01-LM010101.



\section*{Bibliography}
\bibliographystyle{elsarticle-num} 
\bibliography{references}






\end{document}